\documentclass[aps,preprint,amsmath,amssymb]{revtex4-1}
\usepackage{axodraw}
\usepackage{graphicx}
\newcommand{\nn}{\nonumber}
\newcommand{\bd}{\begin{document}}
\newcommand{\ed}{\end{document}}
\newcommand{\bc}{\begin{center}}
\newcommand{\ec}{\end{center}}
\newcommand{\be}{\begin{eqnarray}}
\newcommand{\ee}{\end{eqnarray}}
\newcommand{\ba}{\begin{array}}
\newcommand{\ea}{\ed{array}}
\newcommand{\strich}[1]{#1  \! \! \slash}
\newcommand{\eqn}{\global\def\theequation}
\newcommand{\sw}{sin^2 \theta_W}
\newcommand{\fbd}{f_B}
\renewcommand{\thefootnote}{\alph{footnote}}
\newcommand{\se}{\section}
\newcommand{\sse}{\subsection}
\newcommand{\bi}{\bibitem}
\def\figcap{\section*{Figure Captions\markboth
     {FIGURECAPTIONS}{FIGURECAPTIONS}}\list
     {Figure \arabic{enumi}:\hfill}{\settowidth\labelwidth{Figure 999:}
     \leftmargin\labelwidth
     \advance\leftmargin\labelsep\usecounter{enumi}}}
\let\endfigcap\endlist \relax
\def\reflist{\section*{References\markboth
     {REFLIST}{REFLIST}}\list
     {[\arabic{enumi}]\hfill}{\settowidth\labelwidth{[999]}
     \leftmargin\labelwidth
     \advance\leftmargin\labelsep\usecounter{enumi}}}
\let\endreflist\endlist \relax

\begin{document}
\title
{\Large {\bf Study of $\pi^{0}$ and $\eta$ decays containing dilepton }
}

\author{ \bf \large Chong-Chung Lih$^{1,2}$\\}

\affiliation{
$^{1}$Department of Optometry, Shu-Zen College of Medicine and Management,
Kaohsiung Hsien 452,Taiwan   \\
$^{2}$Department of Physics, National Tsing-Hua University,
Hsinchu 300, Taiwan}

\date{\today}

\begin{abstract}

We calculate
the momentum dependent form factors of $M \to \gamma^* \gamma^*$($M=\pi^{0}, \eta$)
within the light-front quark model.
Using the form factors, we examine the decays of $M \to l^+ l^-$,
$M \to l^+ l^- \gamma$ and $M \to l^+ l^- l^+ l^-$($l=e$ or $\mu$) and
compare our results with the experimental data and other theoretical predictions.
In particular, for $\pi^0 \to e^+e^-$, we find that
the decay branching ratio is
$6.68\times 10^{-8}$, which is closed to the recent measurement of
$(7.48\pm0.29\pm0.25)\times 10^{-8}$ by  E799 of KTeV/Fermilab.

\end{abstract}

\maketitle %

\se{Introduction}

The neutral pseudoscalar meson decays of $M \to l^+ l^-$, in particular $K_L \to \mu^+ \mu^-$,
have played very important roles to understand
the Standard Model (SM). For the light pseudoscalar mesons of $\pi^0$ and $\eta$,
the decays are dominated by the long distance (LD)
contributions, described by the two photon
intermediate state at the lowest order of QED.
Since the short distance (SD) contributions in the SM are many orders of magnitude
smaller, they can be neglected.
Therefore, these decay modes are good processes to explore new physics beyend the SM.

The measurement on this process by the KTeV-E799
experiment at Fermilab has given\cite{ex1}
\be
{\cal B}(\pi^{0} \to e^+ e^-,\, x_D > 0.95) = (6.44\pm0.25\pm0.22)\times 10^{-8}
\ee
where $x_D\equiv (m_{2e}/m_{\pi})^2$ is the Dalitz variable with $m_{2e}$ being the $e^+e^-$ mass.
By extrapolating the Dalitz branching ratio to the full range of $x_D$ with
the overall radiative correction, one gets
\be
{\cal B}^{KTeV}_{\pi^{0} \to e^+ e^-} = (7.48\pm0.29\pm0.25)\times 10^{-8}\,.
\label{KTeV}
\ee
The decay of $\pi^{0} \to e^+ e^-$ has been well studied theoretically over the years.
However,
the KTeV result in Eq. (\ref{KTeV})
 disagrees with the
some theoretical predictions about 1.5 $\sim$ 3.3
standard deviations \cite{ex2,chpt,chpt2,vmd,qm,qed}.

At the lowest order of QED, the
decay branching ratio of $\pi^{0} \to e^+\,e^-$ is found to be\cite{im1,im2,im3}:
\be
{\cal B}_{\pi^{0} \to e^+ e^-}\equiv{\Gamma(\pi^{0}\to e^+ e^-)\over{\Gamma(\pi^{0}
\to2 \gamma)}}= 2\beta \bigg({\alpha\,m_e
\over{\pi m_{\pi}}}\bigg)^{2}\,
|\,{\cal A}(m_{\pi}^{2})|^2,
\label{brall}
\ee
where $\beta\equiv \sqrt{1-4m^2_e/m^2_{\pi}}$
and $|\,{\cal A}(m_{\pi}^{2})|^2$ can be
generally decomposed into $|{\rm Im}\,\,{\cal A}(m_{\pi}^{2})|^2
+|{\rm Re}\,\,{\cal A}(m_{\pi}^{2})|^2$.
Here, ${\rm Im}\,{\cal A}$ denotes the absorptive contribution
from the real photon in the intermediate state, which
can be determined in a model-independent form\cite{im1,im2,im3,im4}
\be
|{\rm Im}\,\,{\cal A}|^2={\pi^2 \over{4\beta^2}}
\,\Bigg[\ln{1-\beta\over{1+\beta}}\Bigg]^2 ,
\label{imaginary}
\ee
leading to the unitary bound on the branching ratio as
\be
{\cal B}_{\pi^{0} \to e^+ e^-} > 2\beta \bigg(\frac{\alpha m_e}
{\pi m_{\pi}}\bigg)^{2}|{\rm Im}\,\,{\cal A}|^2
=4.75 \times 10^{-8}\,\,.
\ee
The real part ${\rm Re}\,{\cal A}$ is given by the dispersive one, which can be
written as the sum of SD and LD contributions,
\be
{\rm Re}\,\,{\cal A}={\rm Re}\,\,{\cal A}_{SD}+{\rm Re}\,\,
{\cal A}_{LD}\,.
 \label{ampsl}
\ee
In the SM, the SD part is given by one-loop box and penguin
diagrams\cite{sd1,sd3}. The LD one involves the form factor
related to the $\pi^{0} \gamma \gamma$ vertex. Using the form factor, the LD amplitude one
has
\be
{\cal A}_{LD}={2i\over{\pi^2 m_{\pi}^2}}\int d^4 q\,{[P^2q^2-(P\cdot q)^2]
\over{q^2\,(P-q)^2\,[(q-p_e)^2-m^2_e]}}\, {F(q^2,(P-q)^2)\over{F(0,0)}}\, ,
\label{Rloop}
\ee
where $P$ and $p_e$ are the pion and electron monenta, respectively. The function
$F(q^2,(P-q)^2)$ is the double form factor of $\pi^0 \to \gamma^*\gamma^*$.
This form factor contains the nontrivial dynamics of the process and has been studied
in various models\cite{qed4e1,qed4e2,vmd4e,chpt,vmd,qm,qed}.
In this paper, we calculate the form factor
$F(q^2,(P-q)^2)$ within the light-front quark model (LFQM)
and use this form factor to evaluate the decays of $\pi^{0} \to e^+ e^-$ and $e^+ e^- \gamma$.
We will also study $\eta$ decays, which  contain a dilepton or dileptons.

This paper is organized as follows:  In Sec.~II, we present
the relevant formulas for the matrix elements and form factors
for $M \to \gamma^* \gamma^*\ (M=\pi^0,\eta)$.
In Sec.~III, we show our numerical results on the form factors and the
branching ratios of meson $M$ decays with dilepton.
We give our conclusions in Sec.~IV.

\se{The form factors}

To calculate $M\to \gamma^*\gamma^*(M=\pi^0, \eta)$
transition from factors within the LFQM,
we have to decompose the mesons into  $Q\bar{Q}$ Fock states. Explicitly, $\pi^0$
may be described as $(u\bar{u}-d\bar{d})/\sqrt{2}$ and the valence state of
$\eta$ can be written as\cite{flavor}
\be
|\eta\rangle=\Phi^{8}\cos\theta_{P}|u\bar{u}+d\bar{d}-2s\bar{s}\rangle/\sqrt{6}-
\Phi^{1}\sin\theta_{P}|u\bar{u}+d\bar{d}+s\bar{s}\rangle/\sqrt{3}\,,
\ee
where $\Phi^{1,8}$ are the wave functions of the Fock states and $\theta_{P}\sim -20^o$
is the mixing angle.
In the scheme of the $Q\bar{Q}$ state, the
amplitude of $M \to \gamma^*\gamma^*$ with $CP$ conservation is given by:
\be A(Q\bar{Q}(P)\to
\gamma^*(q_1,\epsilon_1)~\gamma^*(q_2,\epsilon_2))
=ie^{2}F_{Q\bar{Q}}(q^2_1,q^2_2)~\varepsilon_{\mu\nu\rho\sigma}~\epsilon^\mu_1
 ~\epsilon^\nu_2 ~q^\rho_1 ~q^\sigma_2\,, \label{def}
\ee
where $F_{Q\bar{Q}}(q^2_1,q^2_2)$ in Eq. (\ref{def})
is a symmetric function under the interchange of $q^2_1$ and $q^2_2$.
From the quark-meson diagram depicted in Fig.~1, we get
\begin{figure}[htbp]
\includegraphics*[width=2in,height=6in,angle=-90]{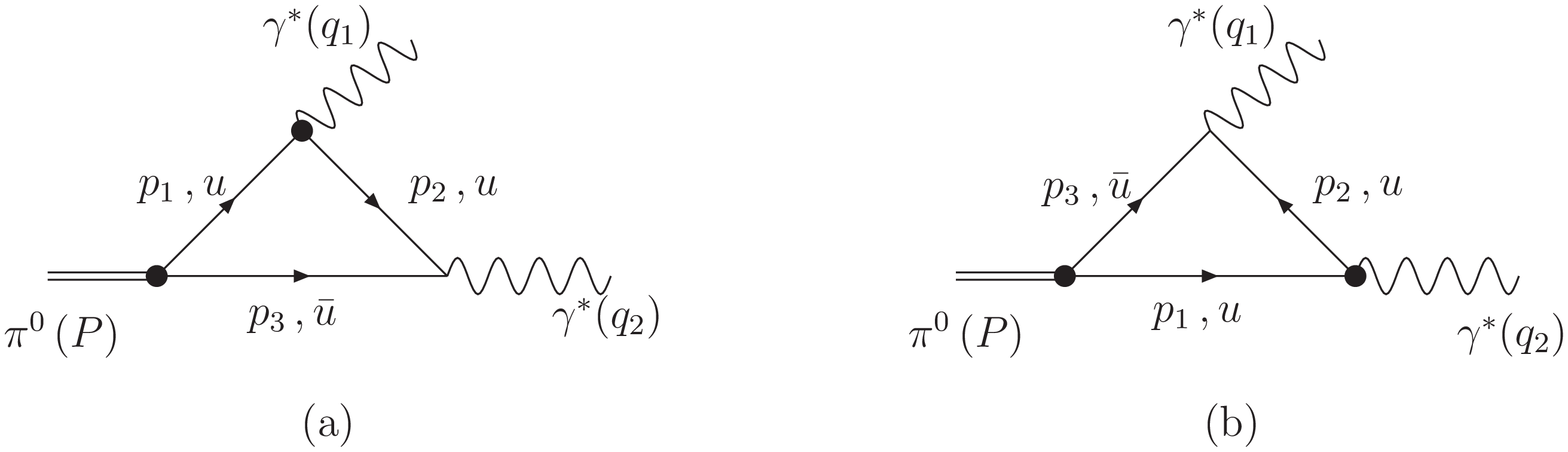}
\caption{  Loop diagrams that contribute of $\pi^{0} \to \gamma^*\gamma^*$.}
\label{F1}
\end{figure}
\be
A(Q\bar{Q}\to \gamma^*(q_1)~\gamma^*(q_2)) &=&
e_{Q}e_{\bar{Q}} N_{c}\int {d^4 p_3 \over{(2 \pi)^4}}
\Lambda_{P}\Bigg\{{\rm Tr}\Bigg[\gamma_5
        {i(-\not{\! p_3}+m_{\bar{Q}})\over{p_3^2-m^2_{\bar{Q}}+i\epsilon}}\not{\! \epsilon_2}
       {i(\not{\!p_2}+m_Q)\over{p_2^2-m^2_Q+i\epsilon}} \nn \\
&&\times \not{\! \epsilon_1}
        {i(\not{\! p_1}+m_Q)\over{p_1^2-m^2_Q+i\epsilon}}
\Bigg]+(\epsilon_1 \leftrightarrow \epsilon_2\,,\,q_1 \leftrightarrow q_2) \Bigg\}
 \nn \\
&&+(\,p_{1(3)} \leftrightarrow p_{3(1)}\,,\, m_{Q} \leftrightarrow
m_{\bar{Q}}) \,,
\label{matrix}
\ee
where $N_c$ is the number of
colors and $\Lambda_{P}$ is a vertex function which related to the
$Q\bar{Q}$ meson. In the light front (LF) approach, the
LF meson wave function
can be expressed by an anti-quark $\bar{Q}$ and a quark
$Q$ with the total momentum $P$  as:
\begin{eqnarray}
|M (P,S,S_z)\,\rangle&=& \sum_{\lambda _{1}\lambda_{2}}\int
[dp_{1}][dp_{2}]
2(2\pi)^{3}\delta ^{3}(P-p_{1}-p_{2})  \nonumber \\
&& ~~~~~~~~ \times \Phi_{M}^{SS_z}(z,k_{\bot}) b_{\bar{Q}}^{+}(p_{1},\lambda _{1}) d_{Q}^{+}(
p_{2},\lambda _{2}) |0\,\rangle\,,
\end{eqnarray}
and
\be
 [d^3p] = {dp^+d^{2}p_{\bot}\over 2(2\pi)^3 }\,,
\ee
where $\Phi_{M}^{\lambda _{1}\lambda _{2}}$ is the amplitude of
the corresponding $\bar{Q}(Q)$ and $p_{1(2)}$ is the on-mass shell
LF momentum of the internal quark.
In the momentum space, the wave function $\Phi_{M}^{SS_z}$ is given by
\be
\Phi_{M}^{SS_z}(k_1,k_2,\lambda_1,\lambda_2)
= R^{SS_z}_{\lambda_1\lambda_2}(z,k_\bot)~ \phi(z, k_\bot),
\label{phi1}
\ee
where $\phi(z,k_{\bot})$ describes the momentum distribution
amplitude of the constituents in the bound state and
$R^{SS_z}_{\lambda_1\lambda_2}$ constructs a spin state $(S,S_z)$
out of light front helicity eigenstates $(\lambda_1\lambda_2)$\cite{melosh}.
The LF relative momentum
variables $(z,k_{\bot})$ are defined by
\be
&& p^+_1=z P^+, \quad p^+_2=(1-z) P^+\,,  \nonumber \\
&& p_{1\bot}=z P _\bot-k_\bot, \quad p_{2\bot}=(1-z)
P_\bot+k_\bot\,. \ee
The normalization condition of the meson
state is given by
\be
&&\langle M(P',S',S'_z)|M(P,S,S_z)\rangle =
2(2\pi)^{3} P^{+} \delta^{3}( P'- P)\delta_{S'S}\delta_{S'_zS_z}
\, , \ee
which leads the momentum distribution amplitude
$\phi(z,k_\bot)$ to
\be
N_c \int {dz\, d^2k_\bot\over 2(2\pi)^3}
|\phi(z,k_\bot)|^2 = 1\, .
\ee
We note that Eq.~(\ref{phi1}) can,
in fact, be expressed as a covariant form\cite{vex1,vex2,lf1}
\be
\Phi_{M}^{SS_z}(z,k_{\bot })&=&\left( \frac{%
p_{1}^{+}p_{2}^{+}}{2[M_{0}^{2}-\left( m_{Q}-m_{\bar{Q}} \right) ^{2}]}\right)^{%
\frac{1}{2}}\overline{u}\left( p_{1}, \lambda _{1}\right)
\gamma^{5}v\left( p_{2},\lambda _{2}\right) \phi(z,k_{\bot}) \,,
 \nn \\
M_0^2&=&{ m_{\bar{Q}}^2+k_\bot^2\over z}+{ m_{Q}^2+k_\bot^2\over
1-z}\, .
\label{n6}
\ee
In principle, the momentum distribution
amplitude $\phi(z,k_\bot)$ can be obtained by solving the
light-front QCD bound state equation \cite{lf1}. However, before
such first-principle solutions are available, we would have to be
contented with phenomenological amplitudes. One example that has
been used is the Gaussian type wave function\cite{lf2,lf3,lf4}:
\be
\phi(z,k_{\bot})=N\sqrt{\frac{1}{N_c}\frac{dk_{z}}{dz}} \exp
\left( -\frac{\vec{k}^{2}} {2\omega_{M}^{2}}\right) \,,
\label{7}
\ee
where $N = 4 ( \pi/\omega_{M}^{2})^\frac{3}{4}$, $\vec k =
(k_{\bot}, k_z)$, and $k_z$ defined through
\be
z = {E_1+k_z\over
E_1 + E_2} \,,~~ \ \ 1-z = {E_2-k_z \over E_1 + E_2} \, , ~~\ \
E_i = \sqrt{m_i^2 + \vec k^2} \,
\ee
by
\be
 & &
\ \ k_{z} =\left( z -\frac{1}{2}\right) M_{0}+\frac{m_{\bar{Q}}^{2}-m_{Q}^{2}}{%
2M_{0}}~\,,~~ M_0=E_1 + E_2\, . \ee and $dk_z/ dz = E_1 E_2/ z(1-
z) M_0$. After integrating over $p_3^-$ in Eq.~(\ref{matrix}), we
obtain \be A(Q\bar{Q}\to \gamma^*(q_1)~\gamma^*(q_2))
&=&e_{Q}e_{\bar{Q}} N_{c} \int^{q_{2}^{+}}_{0} dp_{3}^+ \int
{d^{2}p_{3\bot} \over 2(2\pi)^3\prod^3_{i=1} p^+_i} \bigg[
{\Lambda_{P} \over P^--p^-_{1{\rm on }}-p^-_{3{\rm on }}}
(I|_{p^-_3=p^-_{3{\rm on}}}) \nn \\
&& {1 \over q^-_2-p^-_{2{\rm on }}-p^-_{3{\rm on }}}
+(\epsilon_1 \leftrightarrow \epsilon_2,\,q_1 \leftrightarrow q_2) \bigg]
+(p_{1(3)} \leftrightarrow p_{3(1)}) \,,
\label{pole}
\ee
and
\be
I&=&{\rm Tr}[\gamma_5(-\not{\!p_3}+m_{\bar{Q}})\not{\! \epsilon_2}
(\not{\! p_2}+m_Q)\not{\! \epsilon_1}
(\not{\! p_1}+m_Q)]\,,~~~~~~p_{ion}^-={m_i^2+p_{i\bot}^2\over p_i^+}
\label{trace}
\ee
where the subscript $\{on\}$ represents the on-shell particles.
One can extracted the vertex function $\Lambda_{P}$ from
Eqs.~(\ref{matrix}), (\ref{n6}) and (\ref{pole}),
given by \cite{lf6,vex1,vex2}:
\be
\frac{\Lambda_{P}}{{P^--p^-_{1{\rm on }}-p^-_{3{\rm on }}}} &=&
{\sqrt{p_1^{+} p_3^{+}}\over \sqrt{2[M_{0}^{2}-\left( m_{Q}-m_{\bar{Q}} \right) ^{2}]}}\,\phi(z, k_\bot)~\,,
\ee
To calculated the trace $I$, we have
used the definitions of the LF momentum variables
$(z(x),k_{\bot}(k'_{\bot}))$ and taken the frame with the transverse
monentum $(P-q_{2})_{\perp}=0$ for the $Q\bar{Q}$
state($P$) and photon($q_2$) in Fig.~1a. Hence, the relevant
quark variables are: \be
&&p_{1}^{+}=zP^{+},~~p_{3}^{+}=(1-z)P^{+},
~~p_{1\perp}=zP_{{\perp}}-k_\perp,~~p_{3\perp}=(1-z)P_{{\perp}}+k_\perp\,.
\nonumber \\
&&~p_{2}^{+}=xq_{2}^{+}, ~p_{3}^{+}=(1-x)q_{2}^{+},
~p_{2\perp}=xq_{2_{\perp}}-k^{'}_{\perp},
~p_{3\perp}=(1-x)q_{2_{\perp}}+k^{'}_{\perp}\,. \label{transmom}
\ee At the quark loop, it requires that \be
k_\perp=(z-x)q_{2_{\perp}}+k^{'}_{\perp}\,.
\label{momeq}
\ee
The trace $I$ in
Eq.~(\ref{trace}) can be easily carried out. Thus, the form factor
$F(q^2_1,q^2_2)$ in Eq.~(\ref{def}) can be found to be: \be
F_{Q\bar{Q}}(q_{1}^{2},q_{2}^{2}) &=&-8 \sqrt{N_{c}\over 3}
        \int \frac{dx\,d^{2}k_{\bot }}{2\left( 2\pi \right) ^{3}}\Phi
        \left( z,k_{\bot }^{2}\right) {c^{2}_{Q}\over 1-z}
\frac{m_{Q}}{x(1-x)q_{2}^{2}-m_{Q}^{2}-k_{\bot }^{2}}
       +(q_2 \leftrightarrow q_1)  \,,
\label{fffv}
\ee
where $c_{Q}$ is the quark electric charge factor and
\be
\Phi (z,k_{\bot}^2) &=& N \sqrt{ {\frac{z(1-z) }{2 M_0^2}}}
\sqrt{{\frac{dk_{z}}{dz}}}\exp \left( -{\frac{\vec{k}^{2}}{%
2\omega_M^{2}}}\right)\,,  \nn \\
\vec{k}&=&(\vec{k}_{\bot}, \vec{k}_{z}) \,,  ~~
z=xr\,,~~  \nn \\
r&=&\frac{q_{2}^{+}}{P^+}=\frac
{(m_{P}^{2}+q_{2}^{2}-q_{1}^{2})+
\sqrt{(m_{P}^{2}+q_{2}^{2}-q_{1}^{2})^{2}-4q_{2}^2 m_{P}^{2}}}
{2m_{P}^{2}}\, \,.
\label{res}
\ee
If $q_1$ and  $q_2$ are on mass shell where $r=1$,
the form factors of $\pi \to \gamma \gamma$ and $\eta \to \gamma \gamma$ can be written as
\be
F_{\pi \to \gamma \gamma}(0,0)&=&8\sqrt{2}\sqrt{N_{c}\over 3}
        \int \frac{dx\,d^{2}k_{\bot }}{2\left( 2\pi \right) ^{3}} {\Phi
        \left( x,k_{\bot }^{2}\right)\over 1-x}
\left\{ \frac{4}{9}\frac{m_{u}}{m_{u}^{2}+k_{\bot }^{2}}
-\frac{1}{9}\frac{m_{d}}{m_{d}^{2}+k_{\bot }^{2}} \right\} \,,   \nn \\
F_{\eta \to \gamma \gamma}(0,0)&=&16\sqrt{N_{c}\over 3}
        \int \frac{dx\,d^{2}k_{\bot }}{2\left( 2\pi \right) ^{3}}
\bigg\{ {\Phi^{8}\left( x,k_{\bot }^{2}\right)\cos\theta_{P}\over (1-x)\sqrt6}
\bigg(\frac{4}{9}\frac{m_{u}}{m_{u}^{2}+k_{\bot }^{2}}
+\frac{1}{9}\frac{m_{d}}{m_{d}^{2}+k_{\bot }^{2}}
-\frac{2}{9}\frac{m_{s}}{m_{s}^{2}+k_{\bot }^{2}}\bigg)  \nn \\
&&-{\Phi^{1}\left( x,k_{\bot }^{2}\right)\sin\theta_{P}\over (1-x)\sqrt3}\bigg(
\frac{4}{9}\frac{m_{u}}{m_{u}^{2}+k_{\bot }^{2}}
+\frac{1}{9}\frac{m_{d}}{m_{d}^{2}+k_{\bot }^{2}}
+\frac{1}{9}\frac{m_{s}}{m_{s}^{2}+k_{\bot }^{2}} \bigg)\bigg\} \,.
\label{realff}
\ee

\se{Numerical Result}

To numerically calculate the transition form factors of $\pi^0$
and $\eta$ in Eq.(\ref{fffv}) and (\ref{realff}), we need to
specify the parameters appearing in $\phi(x,k_\bot)$. To
constrain the quark masses of $m_{u,d,s}$ and the meson scale
parameters of $\omega_{M}$ in Eq. (\ref{fffv}), we use the meson decay
constants $f_M$ and its branching ratios of $M \to 2\gamma$,
given by\cite{pdg}
\be
f_{\pi^0}&=&\,132\,{\rm
MeV},~~f_\eta^8=\,169\,{\rm MeV}\,,~~f_\eta^1=\,145\,{\rm MeV}\,.
\label{fpl}
\ee
and
\be Br_{\pi^0 \to
2\gamma}=\,(98.832\pm0.034)\%\,,~~Br_{\eta \to
2\gamma}=\,(39.30\pm0.2)\%\,\,, \label{br2r}
\ee
respectively. Here, the explicit expression of $f_M$ is given by\cite{fp}
\be
f_M&=&\,4{\sqrt{N_c}\over\sqrt{2}}\int {dx\,d^2k_\perp\over 2(2\pi)^3}\,\phi(x,
k_\perp)\,{m\over\sqrt{m^2+k_\perp^2}}\,.
\ee
From
\be
{\cal B}_{M \to 2\gamma}&=& \frac{(4\pi \alpha)^{2}
}{64\pi \Gamma_{P}} m_{P}^{3} |F(0,0)_{P \to 2\gamma}|^2 \,,
\ee
we find that $|F(0,0)_{\pi^{0} (\eta) \to 2\gamma}|=0.274(0.272)$
in $GeV^{-1}$. As an illustration, we extracte $m_u=m_d=0.24$,
$m_s=0.38$ and $\omega_{\pi}=0.33$, $\omega_{\eta1}=0.42$,
$\omega_{\eta8}=0.58$ in GeV, which will be used in our following
numerical calculations.

\sse{$\pi^{0}(\eta) \to e^+ e^- \gamma$}

We now examine process of $\pi^{0} \to e^+e^-\gamma$
with the form factor in Eq.(\ref{fffv}).
The interaction between the photon and leptons is given by the
conventional QED\cite{qed4e1,cqed}. One easily obtains the differential decay rate
\be
{d\,\Gamma(\pi^{0} \to
e^+ e^-\,\gamma)
\over{\Gamma(\pi^{0} \to \gamma\gamma)\,dq^2_1}}=\frac{2\,\alpha}
{3\,\pi}\frac{1}{q_1^2}\,\left(1-\frac{q_1^2}{m_{\pi}^{2}}\right)^{3}
\,\left(1-{4\,m^2_e\over{q_1^2}}\right)^{1/2}\left(1+{2\,m^2_e
\over{q_1^2}}\right)\,|f(t)|^2\,,
\label{llr}
\ee
where $f(t)=F_{\pi}(q_{1}^{2},0)/F_{\pi}(0,0)$ and $t=q_{1}^{2}/m_{\pi}^{2}$.
Obviously, the branching ratio of $\pi^{0}\to e^+e^-\gamma$ in the Eq.(\ref{llr})
depends on the factor of $1/q_{1}^{2}$.
The function of $f(t)$ is an analytic function in the entire physics region
of $4m^2_e \leq q_{1}^{2}\leq m_{\pi}^{2}$, related to
\be
F_{\pi}(q_{1}^{2},0)&=&-4\sqrt{2}
        \int \frac{dx\,d^{2}k_{\bot }}{2\left( 2\pi \right) ^{3}}\Phi
        \left( z,k_{\bot }^{2}\right) {1\over 1-z} \nn\\
&&\bigg\{\frac{4}{9}\bigg[ \frac{m_{u}}{x(1-x)q_{1}^{2}-m_{u}^{2}-k_{\bot }^{2}}
        +\frac{m_{u}}{m_{u}^{2}+k_{\bot }^{2}}\bigg]
-\frac{1}{9}(m_u \leftrightarrow m_d)\bigg\}  \,.
\ee
Integrating over $q^2_1$ in Eq. (\ref{llr}), we
obtain the branching ratio
\be
{\Gamma(\pi^{0}\to e^+e^-\gamma)
\over{\Gamma(\pi^{0}\to\gamma \gamma)}}=1.18 \times 10^{-2}\,,
\ee
which agrees well with those by QED\cite{qed4e1,qed4e2}
and vector meson dominance(VMD) model\cite{vmd4e}.
Our result is also close the experimental data:
${\cal B}_{\pi^0 \to e^+e^-\gamma}^{exp}=(1.198 \pm 0.032)\times 10^{-2}$ \cite{pdg}.

Similarly, the branching ratios of $\eta \to e^+ e^- \gamma$ and $\eta \to \mu^+ \mu^- \gamma$
which normalized with $\eta$ tatal width are found to be
\be
{\cal B}_{\eta \to e^+e^-\gamma}&=&{\Gamma(\eta\to e^+e^-\gamma)
\over{\Gamma_\eta}}=6.95 \times 10^{-3}\,,  \nn\\
{\cal B}_{\eta \to \mu^+ \mu^-\gamma}&=&{\Gamma(\eta\to \mu^+ \mu^-\gamma)
\over{\Gamma_\eta}}=2.94 \times 10^{-4}\,.
\ee
Ours result of $\eta \to e^+ e^- \gamma$ is smaller than
that in the CLEO data\cite{cleo} but
larger than the one in Ref.\cite{mpp}.
However, for the mode of $\eta \to \mu^+ \mu^- \gamma$,
our result agrees with Ref.\cite{mpp} as well as that by the
effective mass theory(EMT)\cite{emt}.
Furthermore, our predictions in the two decay modes agree well
with the experimental data in CELSIUS\cite{wasa} and the PDG\cite{pdg}.

\sse{$\pi^{0}\to e^+e^-e^+e^-$ and $\eta\to \ell^+\ell^-\ell^+\ell^-\ (\ell=e,\mu)$}

We examine the
double lepton-pair decay of $\pi^{0} \to e^+e^-e^+e^-$
with the form factors in Eq.~(\ref{fffv}).
The decay matrix element is calculated
by the conventional QED with the interaction
of $\pi^{0}$ and two photons and
the differential decay rate is given by
\be
{d\,\Gamma(\pi^{0} \to e^+e^-e^+e^-)\over{\Gamma(\pi^{0}\to
\gamma\gamma)\,dq_1^2\,dq_2^2}}={2\over{q_1^2q_2^2}}
\left({\alpha\over{3\pi}}\right)^2\left|{F_{\pi}(q_1^2,q_2^2)
\over{F_{\pi}(0,0)}}\right|^2\,\lambda^{3/2} \left(1,{q_1^2
\over{m^2_{\pi}}},{q_2^2\over{m^2_{\pi}}}\right)\,G_l(q_1^2)\,G_{l'}(q_2^2).
\label{4l}
\ee
where
\be
\lambda (a,b,c)=a^2+b^2+c^2-2(ab+bc+ca),  \nn \\
G_l(q^2)=\left(1-{4\,m^2_e\over{q^2}}\right)^{1/2}\left(1+{2\,m^2_e
\over{q^2}}\right)
\ee
After the integrations over $q^2_1$ and $q^2_2$,
we obtain the branching ratio as follows:
\be
{\cal B}_{\pi^{0}\to e^+e^-e^+e^-}&\equiv&{\Gamma(\pi^{0}\to e^+e^-e^+e^-)
\over{\Gamma(\pi^{0}\to\gamma \gamma)}}= 3.29
\times 10^{-5}\,,
\label{Br4l}
\ee
which is smaller than that in Ref.\cite{qed4e1}, but
larger than the one in Ref.\cite{qed4e2} slightly.
However, all results are consistent with the experimental data.
We note that even if the form factor is replaced by an on-shell constant with
$F(q_{1}^{2},q_{2}^{2})=F(0,0)$, the branching ratio is found to be
very close to the result in Eq.~(\ref{Br4l}).
It might be a good approximation to neglect the momentum dependence of the form factor for the
decay.

We can also perform the similar calculations for $\eta\to l^+l^-l^+l^-$($l=e$ or $\mu$)
and we find
\be
{\cal B}_{\eta \to e^+e^-e^+e^-}&=&2.47 \times 10^{-5}\,,  \nn\\
{\cal B}_{\eta \to e^+e^-\mu^+ \mu^-}&=&5.83 \times 10^{-7}\,,  \nn\\
{\cal B}_{\eta \to \mu^+ \mu^-\mu^+ \mu^-}&=&1.68 \times 10^{-9}\,.
\ee
Our result on ${\cal B}_{\eta \to e^+e^-e^+e^-}$
is in good agreement with the experimental data
${\cal B}_{\eta \to e^+e^-e^+e^-}^{exp}
=(2.7^{+2.1}_{-2.7stat}\pm0.1_{syst})\times 10^{-5}$\cite{wasa} and
Ref.\cite{mpp}. For other modes, currently, our theoretical predictions
are many orders of magnitude smaller than the experimental
upper bounds \cite{pdg,wasa}.

\sse{$\pi^{0}(\eta) \to \ell^+ \ell^-$}

We first calculate the real part of ${\rm Re}\,\,{\cal A}_{LD}$ in
Eq. (\ref{Rloop}) at the pion momentum limit of $P^{2} \to 0$.
At this limit, the relevant form factor of Eq.~(\ref{fffv}),
given by a triangular quark loop, would be simplify to
\be
F(q^{2},q^{2})
&=&-8\sqrt{2}
        \int \frac{dx\,d^{2}k_{\bot }}{2\left( 2\pi \right) ^{3}}\Phi
        \left( z,k_{\bot }^{2}\right) {1\over 1-z} \nn\\
&&\left\{ \frac{4}{9}\frac{m_{u}}{x(1-x)q^{2}-m_{u}^{2}-k_{\bot }^{2}}
        -\frac{1}{9}\frac{m_{d}
        }{x(1-x)q^{2}-m_{d}^{2}-k_{\bot }^{2}}\right\}  \,.
\label{ffpp}
\ee
One could easily find
\be
{\rm Re}\,\,{\cal A}_{LD}(0) \simeq -20.74\,\,.
\ee
The numerical result is in agreement with the most vector meson dominance(VMD) model
at $P^{2} \to 0$.
This implies the equivalence between the VMD
and LFQM descriptions on the form factors of hadrons
with the relevant vector meson
mass of $M_{V} \sim 2m_{u}$ in the VMD.
To illustrate ${\rm Re}\,\,{\cal A}_{LD}(q^2)$ in the range
$-m_{\pi}^{2}\ge q^2 \ge m_{\pi}^{2}$, we use the dispersive framework proposed in
Ref.\cite{qm}. The real part may be written by a once-subtracted dispersion
relation\cite{ex2,qm,cleo2}
\be
{\rm Re}\,\,{\cal A}_{LD}(q^{2})={\rm Re}\,\,{\cal A}(0)+
\frac{q^{2}}{\pi}\,\int_{0}^{\infty} dq'^{2}
\frac{{\rm Im}\,\,{\cal A}(q'^{2})}{(q'^{2}-q^{2})q'^{2}}
\label{osdr}
\ee
Extrapolating from $q^2 =0$ to $m_{\pi}^{2}$,
we find ${\rm Re}\,\,{\cal A}_{LD}(m_{\pi}^{2})=11.18$.
Since the SD part of ${\rm Re}\,{\cal A}_{SD}$ can be neglected,
we get the branching ratio of the real part in Eq.(1) to be
$1.93 \times 10^{-8}$.
The total decay branching ratio is about $6.68 \times 10^{-8}$.
Our prediction is smaller than the experimental value of
${\cal B}^{\rm {KTeV}}_{\pi^{0}\to e^+e^-}=(7.48\pm0.29\pm0.25)\times 10^{-8}$
measured by KTeV. We note that our result is larger than
the values of $(6.41\pm0.19)\times 10^{-8}$ and $6\times 10^{-8}$
calculated in Ref.\cite{vmd,qm} with the VMD and quark model(QM), respectively,
but closed to $(7\pm1)\times 10^{-8}$ in the Chiral Perturbation Theory(ChPT)\cite{chpt}.
It is clear that we provide a method to calculate the form factor
of $\pi^0 \to \gamma^* \gamma^*$ and get a result in
$\pi^0 \to e^+ e^-$ within the LFQM.

The $\eta \to l^+ l^-$ decay can be analyzed in a similar technique as $\pi^{0}\to e^+e^-$.
In the momentum limit $P^{2}\to 0$, we obtained
\be
{\rm Re}\,\,{\cal A}_{(2e)LD}(0) &\simeq& -22.43\,\,,  \nn\\
{\rm Re}\,\,{\cal A}_{(2\mu)LD}(0) &\simeq& -6.48\,\,.
\label{36}
\ee
Form the dispersive integral in Eq.(\ref{osdr}) and Eq.(\ref{36}), one obtains
\be
{\rm Re}\,\,{\cal A}_{(2e)LD}(m_{\eta}^{2}) &\simeq& 27.11\,\,,  \nn\\
{\rm Re}\,\,{\cal A}_{(2\mu)LD}(m_{\eta}^{2}) &\simeq& -2.81\,\,.
\ee
The SD contributions to the decays can be still ignored and the total branching ratios
are given by
\be
{\cal B}_{\eta \to e^+e^-}&=&4.47 \times 10^{-9}\,,  \nn\\
{\cal B}_{\eta \to \mu^+ \mu^-}&=&5.47 \times 10^{-6}\,.
\ee
One notes that the value of ${\cal B}_{\eta \to e^+e^-}$
is larger than the CLEO result\cite{ex2}.
For the mode of $\eta \to \mu^+ \mu^-$, it is consistent with the CLEO\cite{ex2}
and VMD results\cite{cpt}. It also agrees with the PDG data of
$5.8\pm0.8\times 10^{-5}$.

We summarized the related experimental and theoretical values of the
decay branching ratios of $\pi^{0} \to e^+ e^-\gamma$,
$\pi^{0}\to e^+e^-e^+e^-$ and $\pi^{0}\to e^+e^-$ in Table I and
$\eta \to l^+ l^-\gamma$,
$\eta\to l^+l^-l^+l^-$ and $\eta\to l^+l^-$ in Table II.
\begin{table}[hptb]
\caption{Summary of the decays of $\pi^{0}$ with lepton pair.}\label{table:1}
\begin{tabular}{|c||c|c|c|} \hline
~Br~ & Exp. data  & This work & Other models
\\ \hline\hline
$10^{2}~{\cal B}_{e^+e^- \gamma}$
& $ 1.174\pm 0.035$\cite{pdg} & $1.18 $ & $1.18$\cite{qed4e1}\cite{qed4e2}\cite{vmd4e}
\\ \hline
$10^{5}~{\cal B}_{e^+e^-e^+e^-}$
& $3.34\pm0.16$\cite{pdg} & $3.29 $ & $3.28$\cite{qed4e1}, $3.46$\cite{qed4e2}
\\ \hline
$10^{8}~{\cal B}_{e^+e^-}$
 &  $7.48\pm0.29\pm0.25$\cite{ex1,ex2}  & $6.68$ & $7\pm1$\cite{chpt}, $8.3\pm0.4$\cite{chpt2},
$6.41\pm0.19$\cite{vmd}, $6$\cite{qm},
\\
& $6.46\pm0.33$\cite{pdg} &    & $<4.7$\cite{qed}, $6.23\pm0.09$\cite{ex2,cleo2}
\\ \hline
\end{tabular}
\end{table}

\begin{table}[hptb]
\caption{Summary of the decays of $\eta$ with lepton pair.}\label{table:1}
\begin{tabular}{|c||c|c|c|} \hline
~Br~ & Exp. data  & This work & Other models
\\ \hline \hline
$10^{3}~{\cal B}_{e^+e^- \gamma}$
& $7.8\pm0.5_{stat}\pm0.7_{syst}$\cite{wasa} & $6.95$
&  $9.4\pm0.7$\cite{cleo} ,
\\
& $7.0\pm0.7$\cite{pdg}  &      & $6.31-6.46$\cite{mpp}, $6.5$\cite{emt}
\\ \hline
$10^{4}~{\cal B}_{\mu^+\mu^- \gamma}$
& $3.1\pm0.4$\cite{pdg} & $6.95$
& $2.14-3.01$\cite{mpp}, $3.0$\cite{emt}
 \\ \hline
$10^{5}~{\cal B}_{e^+ e^+ e^- e^-}$
& $2.7^{+2.1}_{-2.7stat}\pm0.1_{syst}$\cite{wasa} & $2.47 $ & $2.49-2.62$\cite{mpp}
\\
& $<6.9$\cite{pdg}  &   &
\\ \hline
$10^{7}~{\cal B}_{\mu^+\mu^-e^+e^-}$
& $<1.6\times 10^{3}$\cite{pdg}   & $ 5.83$  & $1.57-2.21$\cite{mpp}
\\ \hline
 $10^{9}~{\cal B}_{\mu^+\mu^-\mu^+\mu^-}$
& $<3.6\times 10^{5}$\cite{pdg}   & $1.68$ &
\\ \hline
$10^{9}~{\cal B}_{e^+e^-}$
& $<2.7\times 10^{4}$\cite{pdg}   & $4.47$ & $13.7$\cite{vmd} , $4.60\pm0.06$\cite{ex2,cleo2}
\\ \hline
$10^{6}~{\cal B}_{\mu^+\mu^-}$
& $5.8\pm0.8$\cite{pdg}  & $5.47$
&  $5.8\pm0.2$\cite{chpt}, $11.4$\cite{vmd}
\\
&   &
&  $5.11\pm0.20$\cite{ex2,cleo2}, $5.2\pm1.2$\cite{cpt}
\\ \hline
\end{tabular}
\end{table}

\se{Conclusions}

\ \ \

We have calculated the form factors of $P \to \gamma^* \gamma^*$($P=\pi^{0}, \eta$)
directly within the LFQM. In our calculations, we have
adopted the Gaussian-type wave function and evaluated
the form factors for the momentum dependences in the energy regions from $q^2 =0$ to $m_{P}^{2}$.
Using the form factors, we have examined
$\pi^0 \to e^+ e^- \gamma$ and $\pi^0 \to e^+ e^- e^+ e^-$ and
shown that our results on the decay branching ratios
agree well with the
experimental data shown in Table. I.
Our predicted values are also close
to those in the QED and VMD models\cite{qed4e1,qed4e2,vmd4e}.
For $\pi^0 \to e^+e^-$, we have found that ${\cal
B}_{\pi^{0} \to e^+e^-}$ is $6.68\times 10^{-8}$, which
agrees with $(7\pm 1)\times 10^{-8}$ in the ChPT
\cite{chpt} but larger than those in Refs.\cite{vmd,qm,qed}.
We have demonstrated that the long-distance
dispersive contribution in this model is possibly small. However,
like other theoretical predictions,
our result for $\pi^{0} \to e^+e^-$ is also slightly smaller
than the experimental data. Clearly, further theoretical
studies as well as more precise experimental data such
as those from the KTeV-E799 experiment at Fermilab on the spectra of the
decays with lepton pair are needed.
About the $\eta$ decays, our results are all consistent with the experimental data.
In particular, the branching ratios of $\eta \to 2e2\mu$, $\eta \to 4\mu$ and $\eta \to 2e$ are
expected to be 4$\sim$5 orders of magnitude lower than the
current experimental upper limits.

\section{Acknowledgments}
This work is supported in part by the National Science Council of
R.O.C. under Contract NSC-97-2112-M-471-002-MY3.

\end{document}